\documentclass{aastex631}

\shorttitle{Oscillatory reconnection}
\shortauthors{Hong et al.}
\graphicspath{{./}{figures/}}

\begin{document}

\title{Oscillatory magnetic reconnection at a coronal bright point}

\author[0000-0002-3804-7395]{Junchao Hong}
\affiliation{Yunnan Observatories, Chinese Academy of Sciences, Kunming 650216, China}
\email{hjcsolar@ynao.ac.cn}

\author{Jiayan Yang} 
 \affiliation{Yunnan Observatories, Chinese Academy of Sciences, Kunming 650216, China}
\email{yjy@ynao.ac.cn}

\author{Yi Bi} 
 \affiliation{Yunnan Observatories, Chinese Academy of Sciences, Kunming 650216, China}
\affiliation{Yunnan Key Laboratory of the Solar physics and Space Science, Kunming 650216}

\author{Bo Yang}
\affiliation{Yunnan Observatories, Chinese Academy of Sciences, Kunming 650216, China}
\affiliation{Yunnan Key Laboratory of the Solar physics and Space Science, Kunming 650216}



\begin{abstract}
Coronal bright points  are typical small-scale coronal brightenings that consist of a bundle of miniature coronal loops. Using the ultra-high-resolution coronal images from the Extreme Ultraviolet Image onboard Solar Obiter, we report the first observational evidence of oscillatory magnetic reconnection at a coronal bright point (CBP). The reconnection is characterised by two bursty phases defined by a reconnection reversal. In the first phase, a current sheet (C1)  is found to form in front of an expanding loop of the bright point. Interestingly,  C1 shorten to a null point during 10 minutes after reaching its maximum length ($\thicksim$2.4Mm). Less than 3 minutes later,  a new current sheet (C2) was clearly seen to grow out from the null point, but along an orthogonal direction relative to C1. C2 reached a maximum length of $\thicksim$4 Mm in ten minutes and then has become short and invisible in the next few minutes as the reconnection has declined. The magnetic reconnection is evidenced by the brightening, plasma flow and temperature increase at the ends of both C1 and C2.  No significant magnetic cancellation or emergence but gradual convergence has occurred during a few hours before the reconnection underneath the CBP. The transition from C1 to C2 suggests the occurrence of coronal oscillatory reconnection with once reconnection reversal, whereby the inflow and outflow regions in the first  phase become the outflow and inflow regions in the second  phase, respectively. It is further found that the oscillatory reconnection could slightly modulate the change in brightness of the coronal bright point.
\end{abstract}



\section{Introduction} \label{sec:intro}

Coronal bright points (CBPs) are small-scale coronal heating objects, showing isolated brightening in coronal X-ray/EUV images \citep{Madjarska2019, Hosseini2021}. They are first discovered as point-like bright features ubiquitously distributed in both coronal holes and quiet-Sun regions by \citet{Vaiana1973}.  They also show as enhanced radio emission and dark absorption feature in He I 10830 \AA\ \citep{Habbal1988}. Most CBPs have a typical diameter less than 30\arcsec with a brighter core ($<$10\arcsec) surrounded by diffused emission and a typical lifetime below 20 hours,  and  no significant difference exists between CBPs in a coronal hole and ones in a quiet region \citep{Golub1974, Golub1976,  Habbal1990, Harvey1993,Alipour2015}. High-resolution observations suggest that a CBP consists of miniature loop structures that connect  opposite polarities of an underlying network magnetic  bipole on photosphere \citep[e.g][]{Zhang2001,Brown2001}. Hence, the geometry of  a CBP strongly depends on the distribution of the photospheric magnetic flux. A CBP would undergo expansion or shrinkage over time, as the emerging, converging, shearing motion, or cancelation of the underlying opposite polarities \citep{Webb1993,Priest1994, Longcope2001, Kwon2012, Huang2012, Nobrega2023, Duan2024}. In addition, \citet{Hong2014} reported that CBPs are the sources of coronal jets and minifilament eruptions associated with these underlying magnetic evolution, emphasising the fact that CBPs are the miniature version of active regions. 

Observations also reveal that the brightness of CBPs can vary significantly on timescales much shorter than the gradual evolution of the underlying  photospheric magnetic fields during their typical lifetimes.  Occasionally, CBPs exhibit impulsive, microflare-like brightenings due to small-scale eruptions \citep{Hong2014, Mou2018, Madjarska2020, Yang2023}. Frequently, CBPs display  lone-lived intensity oscillation  on time scales of minutes to hours \citep[e.g][]{Habbal1981, Strong1992, Ugarte2004, Kariyappa2008,Tian2008, Zhang2012}.  Repeated coronal magnetic reconnections are thought to be the main contributor on the heating as well as the brightness variation in CBPs \citep{Priest1994, Longcope2001, Zhang2012, Wyper2018}.  However,  magnetic reconnections in CBPs are  difficult to be resolved because of their small scale and compact emission.  It is still needed to study the detail  reconnection process and its impact on CBPs.

 A special  reconnection process, named ``oscillatory reconnection'' , rarely observed in solar atmosphere but conducted by a number of numerical studies \citep[e.g.][]{Craig1991, Murray2009,  McLauhlin2012, Karampelas2022a, Karampelas2023, Tabaot2024}. The oscillatory reconnection  is characterised by cycles of a series of reconnection reversals occurring probably in self manner, where the inflow and outflow magnetic fields of one phase of reconnection become the outflow and inflow fields in the following phase of reconnection, respectively \citep{Murray2009}.  \cite{Karampelas2023} presented that the period for oscillatory reconnection is fully dependent on the properties of the background plasma and magnetic fields near the reconnection site and independent of ambient disturbance.  These simulations  indicate that the oscillatory reconnection is a local self-oscillatory process appearing as successive reconnection reversals in the vicinity of a magnetic null point. The oscillatory reconnection thus leads to the cyclic release of  magnetic energy, so that it  is often proposed to account for a number of phenomena like  quasi-periodic pulsations (QPPs) of flares on the Sun and stellar bodies \citep[e.g.][]{ Li2020,Kupriyanova2020}, quasi-periodic flows or intensity oscillations  associated with  jets \citep[e.g.][]{Hong2019, Joshi2020},  and quasi-periodic fast propagating waves \citep[e.g.][]{Shen2022,Zhou2024}.   Observation searching for oscillatory reconnection  is still necessary  to prove its existence in real life plasma environments. 

In this paper, we report the oscillatory reconnection with a reconnection reversal occurring at a CBP and explore the relationship between the oscillatory reconnection and the brightening of the CBP.  The CBP is detected simultaneously by the near-earth space mission Solar Dynamics Observatory ({\it SDO}; \citealt{Pesnell2011}) and the near-Sun mission Solar Obiter ({\it SolO}; \citealt{Muller2020}).  The CBP is shown as an isolated brightening in the the quiet-Sun region by the  Atmospheric Imaging Assembly (AIA A; \citealt{Lemen2012}) onboard {\it SDO} and is associated with a magnetic bipole indicated by the Helioseismic an Magnetic Imager (HMI; \citealt{Schou2012}) onboard {\it SDO}. It demonstrates a bundle of loop structures interacting with the ambient background field revealed by the Extreme Ultraviolet Image (EUI; \citealt{Rochus2020})  onboard {\it SolO}. With the excellent  {\it SolO} data and {\it SDO} data, we suggest this interaction linking to the process of the oscillatory reconnection at the CBP.

\section{Instruments and overview of the CBP }
On 10 April 2023, {\it SolO} was flying at 0.293 AU away from the Sun and  about 62.6 degrees in advance of the Earth along the ecliptic (More orbit information can be checked from \url{https://stereo-ssc.nascom.nasa.gov/cgi-bin/make_where_gif}), which provided an opportunity to capture the Sun with ultra-high resolution and good signal-to-noise. 
The Extreme Ultraviolet High Resolution Imager ({HRI$_E$$_U$$_V$}) of  {\it SolO}/EUI  focused on a quiet-Sun region (QS) from 03:35 UT to 09:35 UT with  a  field of view (FOV)  $\thicksim$16.8\arcmin$\times$16.8\arcmin , where our interest CBP located. It  obtained 2160 frames of 174  \AA\ images  with a cadence of 10 s and a pixel size of 0.492\arcsec ( $\thicksim$105$Km$). The 174 \AA\ channel is sensitive to coronal plasma at temperatures of  $\thicksim$1 Mk because of the contribution of the emission lines Fe IX  at 171.1 \AA\ and Fe X at 174.5, 177.2 \AA.  These 174 \AA\  images was processed into a sequence of level 2 data  which is easily accessible from \url{https://www.sidc.be/EUI/data/L2/2023/04/10/}. We further removed the remaining spacecraft jitter base on a cross-correlation algorithm and corrected the observation time in line with the observations near Earth. 

Meanwhile, the {\it SDO} captured the CBP from the different viewpoint  in orbit around the Earth. It provided full-disk AIA coronal images  and HMI photospheric line-of-sight magnetograms. The AIA has a cadence of 12 s and a pixels size of 0.6\arcsec ( $\thicksim$440$Km$).    The CBP  stands out clearly in AIA EUV channels like 171 \AA\ and 193 \AA\ .  The AIA 171 \AA\ channel responses slight cooler plasma (Fe IX at $\thicksim$0.7 Mk) than the ({HRI$_E$$_U$$_V$) 174 \AA\, while the 193 \AA\ channel has a higher temperature response  at  $\thicksim$1.6 Mk (dominated by Fe XII) \citep{Odwyer2010}.  The calibrated HMI magnetograms have the same pixel size as the AIA but with a cadence of 45 s, which are used to explore the photospheric magnetic field related to the CBP.

Using the standard reprojection routine {\it reprojetct\_to()} in Sunpy, we further reprojected the   {\it SolO} images to the view from {\it SDO} so that we can explore the 174 \AA\  CBP in alignment with the AIA observations and HMI magnetograms.  Note that whether the projection routine works well on pixels away from the solar centre is dependent of the set solar radius at which emissions in the projected map are assumed to have come from. The default value for the radius is the photospheric radius. However, the CBP should have its EUV emission at heights of several megameters above the photosphere \citep{Kwon2012}. To align the 174 \AA\ CBP with the AIA counterpart one exactly,  we ultimately set the radius as the photospheric radius plus a height of  $\thicksim$6.8 Mm for the performed  {\it reprojetct\_to()} procedure  after multiple experiments with different heights.

\section{Result} \label{sec:style}
\subsection{overview of the CBP}
Figure 1 shows the snapshots of the AIA 193 \AA\  and {HRI$_E$$_U$$_V$} 174 \AA\ on 10 April 2023.  Where the {HRI$_E$$_U$$_V$} 174 \AA\ focused is indicated by the warped blue box in the full-disk AIA image. The green circle in each image marks our interest CBP in the quiet Sun. The CBP appears as a QS point-like brightening in the full-disk 193 \AA\ image while it resides an area with a spatial scale of $\thicksim$10 Mm in the high-resolution 174 \AA\ image.

\subsection{ reconnection reversal in the close-up views of the CBP}
 Figure 2 presents the projected 174 \AA\ images together with the corresponding AIA images and HMI magnetograms.
By projecting the  {HRI$_E$$_U$$_V$} 174 \AA\ images into the view from {\it SDO}, we compare the close-up views of the CBP region watched by the different telescopes at the two moments of 04:09 UT and 04:28 UT in Figure 2.  The main body of the CBP appears as an arcade of  loops connecting a pair of positive/negative polarities P/N, demonstrated by the AIA 193 \AA\ and 171 \AA\ images with the overlaid HMI magnetograms. However, the  {HRI$_E$$_U$$_V$} 174 \AA\ images resolve more outer structures besides the main inner loops. It is distinctly seen that there is a core-elongated  X-shaped bright structure wrapping the west sheath of the 174 \AA\ CBP  at 04:09 UT.   The core of the ``X" geometry is elongated as a line-like structure, highlighted by the cyan line segment C1 at 04:09 in Fig.2. The other similar X-shaped structure is also found at 04:28 UT, featured by the line C2 in the core of ``X".  The bifurcate legs of each X structure are highlighted by the dashed lines in Fig.2, extending away from each end of C1/C2.  

In contrast, these X-shaped structures are almost invisible or display as very  diffuse haze in  the AIA 193 \AA\  and 171 \AA\ images.  By overlying the 174 \AA\ X-shaped  skeletons  on the HMI magnetograms,  we infer that the four feet of each X structure correspond to a quadrupole magnetic configuration. Two feet anchor to the dipole P/N (unsigned magnetic flux densities: 25-100 G) of  the CBP while another two feet reside at the surrounding weak positive/negative polarities (below 25 G).   The observed X-shaped interconnections  are very likely to reflect the magnetic separatrix between different magnetic flux domains associated with quadrupole configuration, where magnetic reconnection prefer to occur  \citep{Liu2016}. According to the standard 2-dimensional model of the magnetic reconnection, C1/C2 trace out  the projection of the current sheets, separating the magnetic fields of the CBP loop and the ambient background fields and providing favourable conditions for magnetic reconnection \citep{Parker1957,1958Sweet,Petschek1964, Longcope1998}. The small-scale current sheets C1/C2 present the typical line-shaped morphology with bifurcating ends in accordance with other ones observed in solar low atmosphere \citep[e.g.][]{ Yang2015,Li2021,Yang2024}.  The widths of C1/C2 are estimated to be only 3 pixels ($\thicksim$0.3 Mm), possibly limited by the HRI$_E$$_U$$_V$ instrument resolution \citep{Ding2024}. 
 
Comparing C1 with C2 in Fig.2 carefully, it is further found that C1 is  perpendicular to C2 although they are produced at the same site. This means that the outflow regions at each end of C1 correspond to the inflow regions at each side of C2. Thus, a magnetic reconnection reversal might occur when the reconnection  associated with C1 enters into  the later reconnection with C2.

Figure 3 presents the entire process of the reconnection reversal  in the 174 \AA\ sequence.  The CBP started to slightly expand outward from 03:56 UT, possibly due to the instability of its core field \citep{Hong2014}. There were at least two groups of the CBP small loops rising up or expanding laterally, as indicated by the white and cyan arrows in the first two panels of Fig.2. In particular, the western group loops of the CBP first expanded outward from 3:56 to 4:02UT, then seemed to encounter some obstruction, stopped, and slightly deformed at 4:04 UT. The deformed loops were likely subjected to some kind of squeezing from invisible ambient magnetic fields,  resulting in the formation and growth of the sheet structure C1  with two bifurcate ends from 04:04 UT. As shown in Fig.3,  C1 has grown up from 04:04 UT to 04:09 UT,  but then shorten to null until 04:16 UT.  It reached its maximum length of $\thicksim$2.4 Mm ($\thicksim$23 pixels)  at 04:09:50 UT.  Some bright blobs were seen moving bi-directionally along C1 and the bifurcation legs, which in turn light the cusp ends of C1 and the bifurcation legs (see the panels at 04:11 UT and 04:14 UT or the associated movie ``animation\_fig3..mp4"). These moving plasma blobs indicate the potential presence of the magnetic reconnection at the X-shaped structure \citep{Ni2017,Yang2018, Yan2022, Cai2024}. The magnetic reconnection took place through C1, consuming the inflowing fields of the CBP loop and producing the bi-directional flows. The bi-directional plasma flows are also found in  CBPs by previous works, but the related reconnection sheets can hardly be resolved \citep{Ning2014, Ning2020, Li2022}.

Interestingly, the reconnection did not end as C1 shrank to null, because a new current sheet ``C2"  came out soon at the same site. Some weak, small blobs are also observed along C2 and its bifurcation ends. However,  C2 is not the reproduction of C1 because the orientation of C2 is orthogonal to that of C1.  It is also indicated by the contrast of the dashed line along C1  at  04:14:20 UT to the dashed line along C2 at 04:27:50 UT. The transition from the disappearance of C1 to the appearance of C2 took less than 3 minutes, suggesting two successive reconnection phases  rather than two isolated reconnection events. This phase transition is featured by the reconnection reversal, i.e. the inflow and outflow fields around C1 became the outflow and inflow fields around C2, respectively. This is the direct evidence of the so-called oscillatory reconnection \citep{Murray2009}. 

C2 eventually reached its maximum length of  $\thicksim$4 Mm  at 04:29:30 UT.  During the following tens of minutes, C2 got shorter and fainter but the reconnection brightening remained at one of its end and the outflow region close to the CBP until 04:35 UT. This implied that the magnetic reconnection should be still ongoing even the current sheet became invisible. From the movie, weak brightening around the reconnection location continued after 04:35 UT.  It is not sure whether the oscillatory reconnection started a new reversal with a new weaker reconnection phase, since no current sheet could be detected yet again. Whatever, due to the reconnection reversal from C1 to C2, the reconnected magnetic fields of the CBP that were consumed in the first reconnection phase should be compensated by the outflow fields produced in  the second reconnection phase.  As a result, the CBP basically sustained its initial magnetic configuration, even it underwent two successive impulsive energy releases through the oscillatory reconnection.

Consecutive difference 174 \AA\ images  are shown in the first two rows of Figure 4 to display more clearly for the growth and shrink of C1/C2. As we see, C1 got longer from 4:08 UT to 4:09 UT, but got shorter from 04:09 UT to 04:14 UT. To 04:17 UT, the current sheet shrank to zero. Then,  It is clear than  C2 became longer from 4:27 UT to 4:29 UT but shorter again from 4:29 UT to 4:33 UT.    An associated movie ``animation\_fig4..mp4" is made to show the dynamic evolutions of C1/C2 in the difference images.  Similar in Fig.3, C1 and C2 appeared as two mutually orthogonal  bright linear segments of which the ends are tied with the bifurcation legs. For C1, the 4:14 UT difference images show clearly that both ends of C1 connecting to the CBP footpoints along the left-side legs  and the ambient quiet regions along the right-side legs.  For C2, as shown in the second row,  the bifurcation leg at the left end of C2 connects to the CBP footpoints while  its right end connects to the ambient quiet regions. 

\subsection{photospheric activity and coronal heating associated with the oscillatory reconnection}
The HMI magnetograms in the bottom row of Figure 4 show that there were no considerable emergence and cancellation of the photospheric magnetic fields associated with the CBPs within the time interval before and during the formation of C1/C2. However, the main magnetic polarities (P/N) at the footpoints of CBP have gradually approached to each other during three hours before the  oscillatory reconnection.  The merge of the positive patches around P and fragmentation of the negative patches around N  have also occurred during the three hours. These complicated but slight changes of the quiet-Sun magnetic elements are popular in small-scale events like the evolution of CBPs  and miniature corona mass ejections \citep{Honarbakhsh2016}. 

To track the convergence of the beneath P/N, we calculated the photospheric horizontal flows based on the consecutive HMI magnetograms by using the popular pixel-tracking algorithm (the optical flow method in Python package OpenCV). Figure 5a presents the superimposed AIA 193 \AA\ intensity image and one hour-averaged flow fields on the HMI magnetogram within a larger FOV of the quiet Sun . The red arrows show the direction and magnitude of the movement of the photospheric magnetic elements.  It suggests that the CBP  is located where the general supergranular diverging flows are converging,  consistent with numerous solar small-scale  network activities \citep{Innes2009, Honarbakhsh2016}.  Thus, the network fields below the CBP would be driven to get close.  The zoomed panels (b) and (c) highlight the local flows around the CBPs together with the overlaid 174 \AA\ observation.  It obviously suggests  that the dipole footpoints of the CBP, i.e. the positive (P) and negative (N) polarities, are converging to each other gradually at  a speed less than 0.5 Km/s (about 0.2-0.3 Km/s) during the appearance of C1/C2. This convergence may contribute to the early expansion of the magnetic field of the CBP , which further initiates the reconnection along C1.

The coronal differential emission measure (DEM) analysis was performed to investigate the thermal information on the reconnection based on the six-channels (171,193,211,335,,131,94\AA) of AIA data. We carried the quickly pixel-by-pixel DEM inversions by using the sparse inversion code  \citep{Cheung2015}. The code was set to solve the DEM profile at each pixel with two basic Gaussian functions (with sigmas 0.1 and 0.4), given that most coronal features flux can be best-fitted with the single or double Gaussian DEMs \citep{Aschwanden2011, Hosseini2021}. Figure 6 presents the DEM results for C1 at 04:09 UT and for C2 at 04:33 UT.  The first row shows the maps of EM in the logarithmic temperature range of 6-7, i.e. the spatial distribution of coronal plasmas above million degree. The second row displays the EM-weighted logarithmic temperature maps that give the  spatial distribution of the averaged temperature at each pixel.  The structure of C1/C2 as seen in {HRI$_E$$_U$$_V$}  174 \AA\  remained barely distinguishable in the EM and temperature maps, yet signatures of peripheral emissions and localized heating persisted around the domains of C1/C2. In particular, there were remarkably temperature and emission enhances at the ends of C1/C2, suggesting the heating in the outflow regions of the reconnections along C1/C2.  The DEM profiles in the bottom row show the Gaussian-shaped thermal distribution at the ends of C1/C2, with the average temperatures of 2.4-2.6 Mk. 

\subsection{ Oscillatory of the current sheets lengths  associated with the brightness variation of the CBP }
In order to correlate the oscillatory reconnection with the brightness of the CBP, we estimated the lengths of C1 and C2 defined by the distance between both ends of C1/C2 at different moment, and calculated the {HRI$_E$$_U$$_V$} 174 \AA\ , the AIA 171 \AA\ and 193 \AA\ integrated intensities over the whole CBP. The temporal  evolution of C1 is clearly shown by the  174 \AA\  time-slice plot in Fig.7a, as well as the one of C2 shown in Fig.7b. C1 started to form at about 04:03 UT, reached longest at about 4:09 UT, and became gradually shorter during the last ten minutes of its life. We also tracked both ends of C1/C2 from the 174 \AA\ sequence and overlay their positions (indicated by the pink asterisk) on the time-slice plots. Clearly,  C1 shrank to a null point at about 04:20 UT$\pm$1.5 mins, then C2 started to grow up from the null point. C2 underwent a slow growing stage from 04:20 UT to 04:27 UT, a rapid growing stage from 04:27 UT to 04:30 UT, and a decay stage after 04:30 UT.

Figure 7c shows the temporal variations of the CBP brightness with the light curves in the passbands of  {HRI$_E$$_U$$_V$} 174 \AA\ ,  AIA 171 \AA\ and 193 \AA\ , as well as the time profiles of C1/C2 lengths.  During the reconnection phase with C1, the increase and decrease of the light curves are nearly co-temporal with the elongation and shorten of C1. During the reconnection phase with C2, the light cures also show slightly increase and decrease along with the change of C2. Note that  the first CBP brightening  is most sensitive in the 193 \AA\ and 174 \AA\ light curves but the second  brightening show mainly a intensity increase in 171 \AA\  and  only a small peak in 174 \AA\ . This difference might be caused by a temperature effect,  since the 193 \AA\ and 174 \AA\  passbands sample hotter plasma than 171 \AA\ .  It is possibly that the first phase of the oscillatory reconnection gives stronger heating to the CBP than the second phase.  As a result, the maximum length of C1 is correlated with the peaks of the 193 \AA\ and 174 \AA\ curves, while the maximal C2  corresponds to a local small peak of either 174 \AA\ or 171 \AA\ curves.  The CBP thus underwent a damped brightness oscillation with a period of about 20 minutes.

\section{ Summary and Discussion}
 Thanks to the high resolution and high signal-to-noise of {\it SolO}/EUI  at 0.293 AU away from the Sun, we  first find the observational signature in support of  oscillatory magnetic reconnection at an ordinary CBP. The CBP consists of typical bipolar magnetic loops in quiet-Sun region, where the oscillatory reconnection takes place at the new formed X-shaped interconnection  between  the expanding CBP loops and ambient fields. The X-shaped interconnection is consistent in geometry with those magnetic setups of X-type null points where oscillatory reconnections are simulated out \citep[e.g.][]{Craig1991, Karampelas2022a}.   The observation suggests that the oscillatory reconnection performed as an overall reconnection process with two individual phases. The first phase of the reconnection was initiated when the current sheet C1 was squeezed out at the core of the X-shaped structure.   C1 exhibited an increase and a decrease in length  during the 17 minutes life of the first phase.  The reconnection did not cease as C1 shrinks into a null point , but stepped into the next phase as the new current sheet C2 grew out from the null point within 3 minutes. The second reconnection phase lasted more than 20 minutes and C2 also displayed an increase and a decrease in length. Weak bi-directional plasma blobs were also observed to move out from C1/C2 , causing the brightening along the bifurcate ends of C1/C2.  The most critical evidence of the oscillatory reconnection was the transition from C1 to C2, which suggested once continuous reconnection reversal as the orientation of C2 was orthogonal to that of C1 \citep{Murray2009}.  
 
 Although {\it SDO} detected the CBP simultaneously, the oscillatory current sheets C1/C2 were not be clearly resolved in {\it SDO}/AIA EUV images. This discrepancy may arise from either the limited spatial resolution of {\it SDO}/AIA compared to {\it SolO}/EUI  or differing observational perspectives between the two spacecraft. However, there were  signatures of very  diffuse haze around the reconnection domains outside the CBPs in  the AIA 193 \AA\  and 171 \AA\ images.  DEM analysis suggested that these haze perhaps came from the heated outflows from the oscillatory reconnection. In special, unique enhances in temperature  and  EM maps were found at each end of the current sheet C1/C2. The oscillatory reconnection heated the localized  plasma outside the CBP to 2.4-2.6 MK, close to the upper limit of the determined temperature for CBPs in \citet{Hosseini2021}. The {\it SDO}/HMI magnetograms demonstrated that the photospheric magnetic field beneath the CBP underwent a slow convergence before and during the oscillatory reconnection. No significant magnetic cancellation or emergence led the occurrence of the oscillatory reconnection as some small-scale eruptions \citep[e.g.][]{Honarbakhsh2016, Xu2024}.  The  reconnection at the CBP was perhaps initiated by the  persistent  photospheric magnetic convergence that was driven by the converging  supergranular flows at the network boundary \citep{McIntosh2007}.
 
 The present reconnection reversal is found in the small-scale structure of the CBP, which is ubiquitous in the solar atmosphere.  It is possible that there could be many events of the oscillatory reconnection on the Sun. We also found the similar  reconnection reversal at the base of a coronal jet as reported in \cite{Hong2019}.  As comparison, the reconnection reversal in \cite{Hong2019}  took place at a null point of the fan-spine magnetic configuration,  while  the reconnection  reversal presented here occurred in the X-shaped magnetic configuration that was formed  at the edge of the CBP. Whatever, the peak sizes of the oscillatory current sheets C1/C2 (2.4/4 Mm in length, 0.3 Mm in width)  and the transition duration (within 3 mins) of the reversal in the CBP are comparable with those measurements  in \cite{Hong2019}   ($\thicksim$3 Mm in length and $\thicksim$0.8 Mm in width for the current sheets, 1 mins in transition duration for the reversal).  \cite{Murray2009} found successive reconnection reversals occurring in a self-consistent manner,  as the gas pressure in the outflow region increased over the Lorentz force from the inflow fields during each reconnection phase in their simulations. We still do not determine if the observed reconnection reversals are initiated in a self-consistent manner. It is worth seeking more events for the reconnection reversals to statistically explore the common periods and decay rates of the oscillatory reconnection in the low corona.

Note that not all reconnection reversal suggests the oscillatory reconnection referred here.  In particular, magnetic configuration can undergo long-periodic changes  through  a series of isolated reconnection events,  appearing as a result of a reconnection reversal \citep[e.g][]{ Sterling2001, Goff2007, Zhang2014, Xue2019}. For this type of reversal,  each reconnection episode is a discrete event, separated  by more than the lifetime of the reconnection self from the next reconnection episode. \cite{Zhang2014} called this type of reconnection reversal as reciprocatory magnetic reconnection which should have different driving mechanism from the oscillatory reconnection. The direct signatures of oscillatory reconnection are plausibly presented  in the jet of \cite{Hong2019} and the CBP here, which reported the uninterrupted reconnection reversals  revealed by the periodic change of the  current sheet within several minutes.

We further find that the CBP brightness fluctuation are well synchronised with the periodic change of the lengths of the current sheets C1/C2 during the oscillatory reconnection.    From the multi-wavelength light curves  of the CBP,  a remarkable intensity peak (i.e. strong brightening) appears when C1 reaches its maximum length in the first reconnection phase, while a mild intensity enhancement (weak brightening) occurs around the peak of C2 during the second reconnection phase.   In general,  the reconnection rate is proportional to the length of the current sheet in each reconnection phase  and the peak of reconnection rate reduces after each reconnection reversal in the oscillatory reconnection   \citep{Murray2009}.  As a result, the CBP  underwent a damped brightness oscillation with a period  defined by the time interval (here about 20 mins) between the peak length of C1 and that of C2. Hence,  the fluctuation of the CBP brightness seemed to be modulated by the oscillatory reconnection because of its periodic energy release.  

The brightness oscillations in CBPs have been reported  extensively in previous studies \citep[e.g.][]{Ugarte2004, Kariyappa2008,Tian2008, Zhang2012,Ramsey2023}. \citet{Ugarte2004} presented that the transition region and coronal lines emission of a CBP experienced a damped oscillation with periods peaking around 10 mins. Via power spectrum analysis, \citet{Kariyappa2008} determined the periods of the brightness oscillations among dozens of CBPs which had different emission levels.  Those CBPs were made up of the similar dipole configurations as our CBP.  They  concluded that different CBPs show similar brightness oscillation periods ranging from a few minutes to hours.  This indicates there might be a common heating mechanism leading the brightness oscillations of  CBPs.  Recurrent  small-scale reconnections or wave-modulated mechanism are thought to be responsible for quasi-periodic heating and brightening in CBPs \citep{Kariyappa2011,  Kumar2011, Zhang2012, Samanta2015}. Here, we propose that the oscillatory reconnection is an appropriate candidate mechanism by which the CBP's brightness oscillates with different periods. Three consequences are expected in this mechanism. First, without additional injection and release of magnetic free energy (e.g magnetic converging, shearing, and cancellation on the photosphere beneath CBPs), the self-initiated oscillatory reconnection will produce a damped oscillatory profile with an increase in period. Furthermore, the oscillatory reconnection can give rise to successive brightenings in CBPs but not destroy  the overall magnetic structures of CBPs, since the reconnection reversal can restore the reconnected fields.  Finally, \cite{Chen1999} determined the typical lifetime of the reconnection in the low corona as only tens of minutes considering that  the reconnected fields are piled up in a limited volume to hinder the reconnection. However, the ongoing oscillatory reconnection can naturally extend the lifetime with the reconnection reversal, although each reconnection phase gradually weakens. Recurrent, persistent heating events have been discovered in the small-scale null-point structure with high-resolution observations \citep[e.g.][]{Cheng2023} . Therefore, we expected that the oscillatory reconnection might provide persistent, quasi-periodic heating to CBPs and even smaller features if the oscillatory regime is popular at a small-scale magnetic null point.   More ultra-high resolution observations are required to check how often the oscillatory reconnection takes place in CBPs.
 
\begin{acknowledgments}
The data used here are courtesy of the {\it Solar Obiter} and {\it SDO} science teams.  This work is
supported by the Strategic Priority Research Program of the Chinese Academy of Sciences, Grant No. XDB0560000, the Natural Science Foundation of China under grants under grants 12173084, 12273108, 12273106, and 12203097, the Yunnan Science Foundation of China (202401AT070071), the Yunnan Province XingDian Talent Support Program, the ``Yunnan Revitalization Talent Support Program" Innovation Team Project (202405AS350012), the CAS ``Light of West China" Program, and the Yunnan Key Laboratory of Solar Physics and Space Science (202205AG070009).
\end{acknowledgments}

\begin{figure}[ht!]
\epsscale{1.2}
\plotone{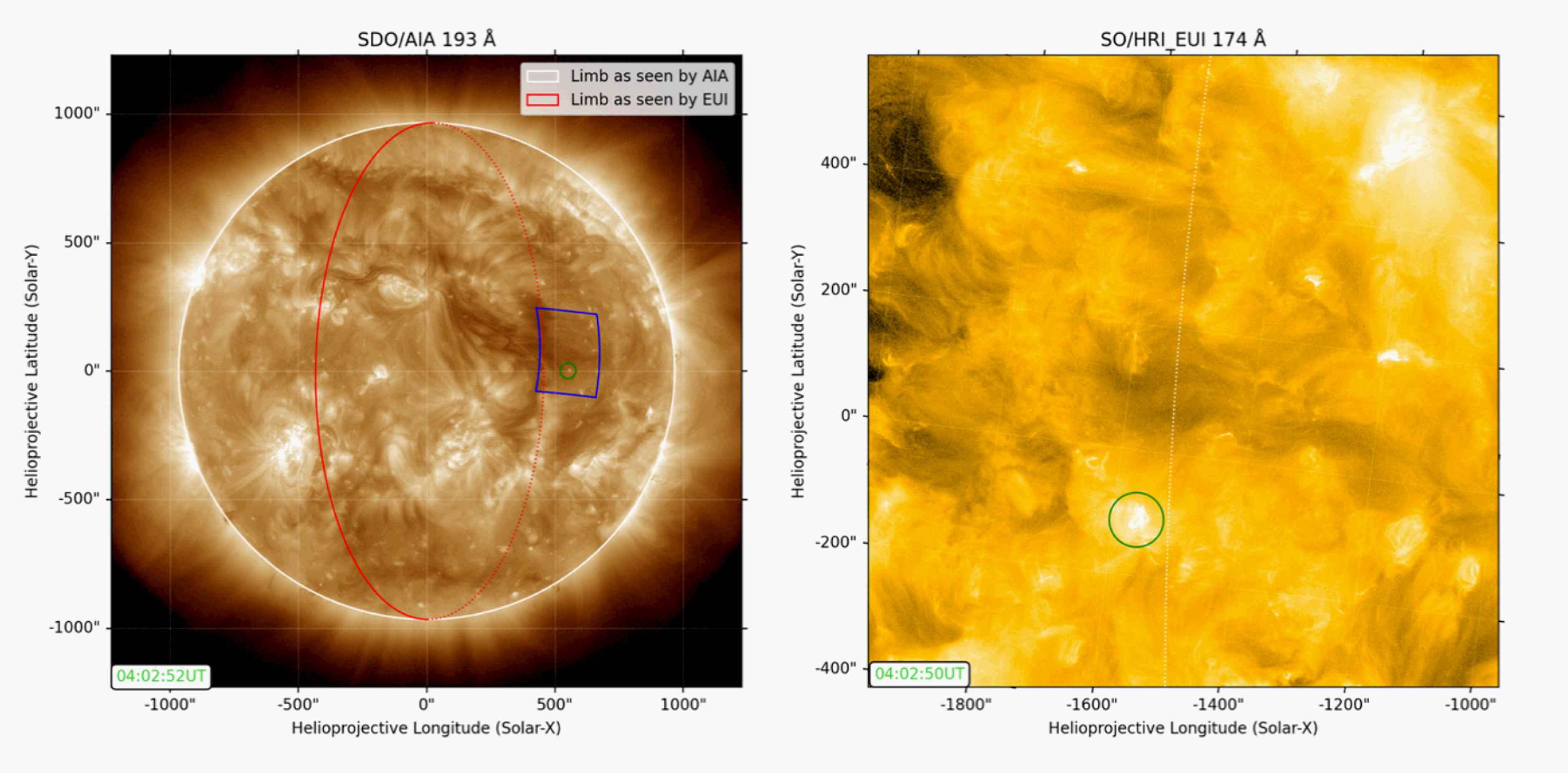}
\caption{ Locations of the targeted CBP observed on 10 April 2023 by SDO/AIA  and Solar Obiter/EUI, respectively. The warped blue box indicates the field of view (FOV) of SO/EUI observations when reprojected into the SDO full-disk coordinated frame.  SoLO was about 62.6 degrees in advance of SDO, indicated by the different solar limbs as seen by them.}
\end{figure}

\begin{figure}[ht!]
\plotone{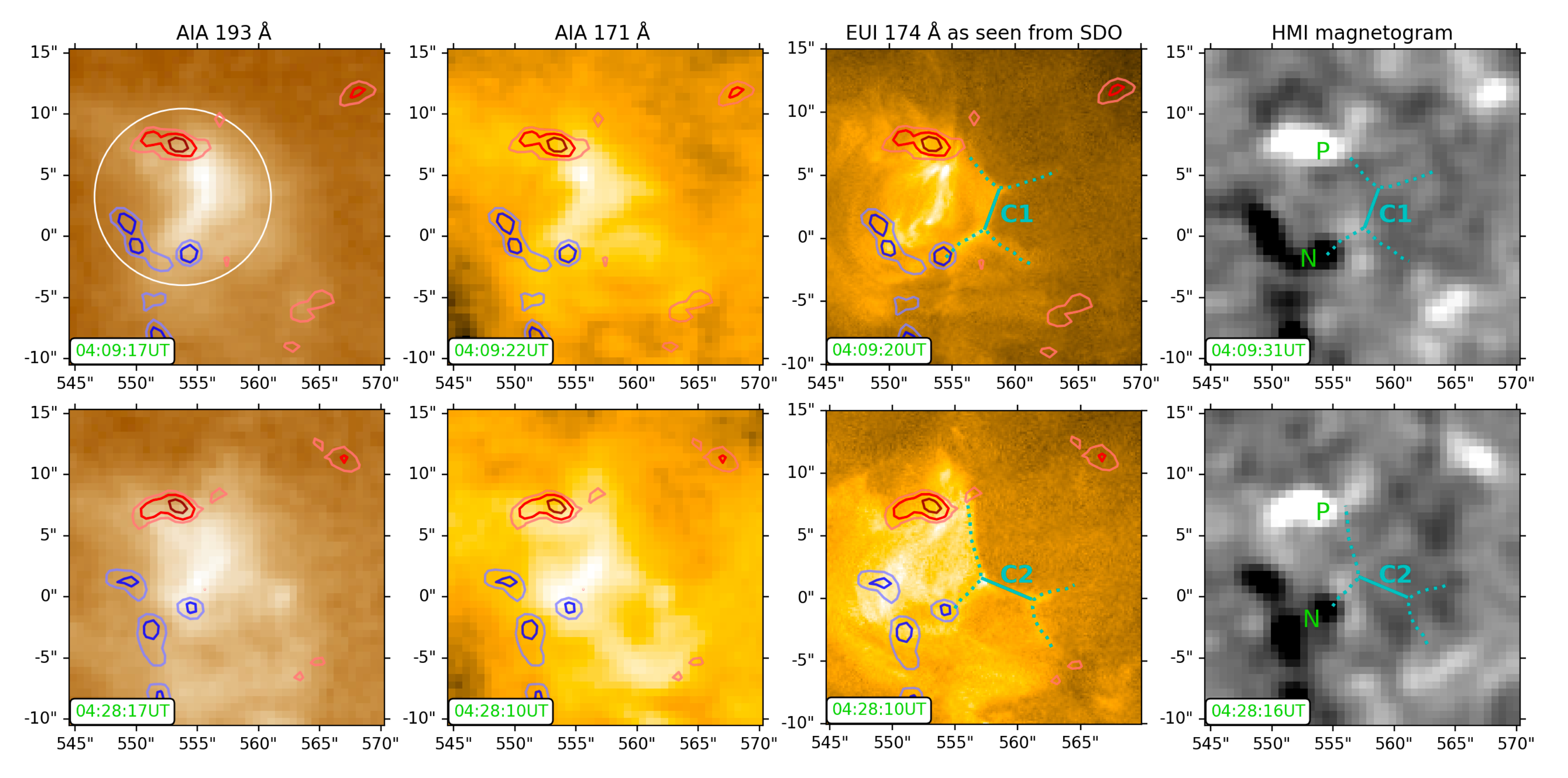}
\caption{Different appearance of the CBP detected by different telescopes.  The  {\it SolO}/EUI 174 \AA\  images are reprojected to the {\it SDO} coordinated frame by using the standard reprojection procedure  in Sunpy. The circle outlines an area where the brightness of the CBP is calculated in Fig.4. The gradient red (blue) contours are the levels of the HMI magnetic field at 25 (-25), 50 (-50), and 100 (-100) G. The cyan lines indicate the current sheets C1/C2 tying the west edge of the CBP. The dashed lines highlight the bifurcation structure connecting two ends of C1/C2.  }
\end{figure}

\begin{figure}[ht!]
\epsscale{1.}
\plotone{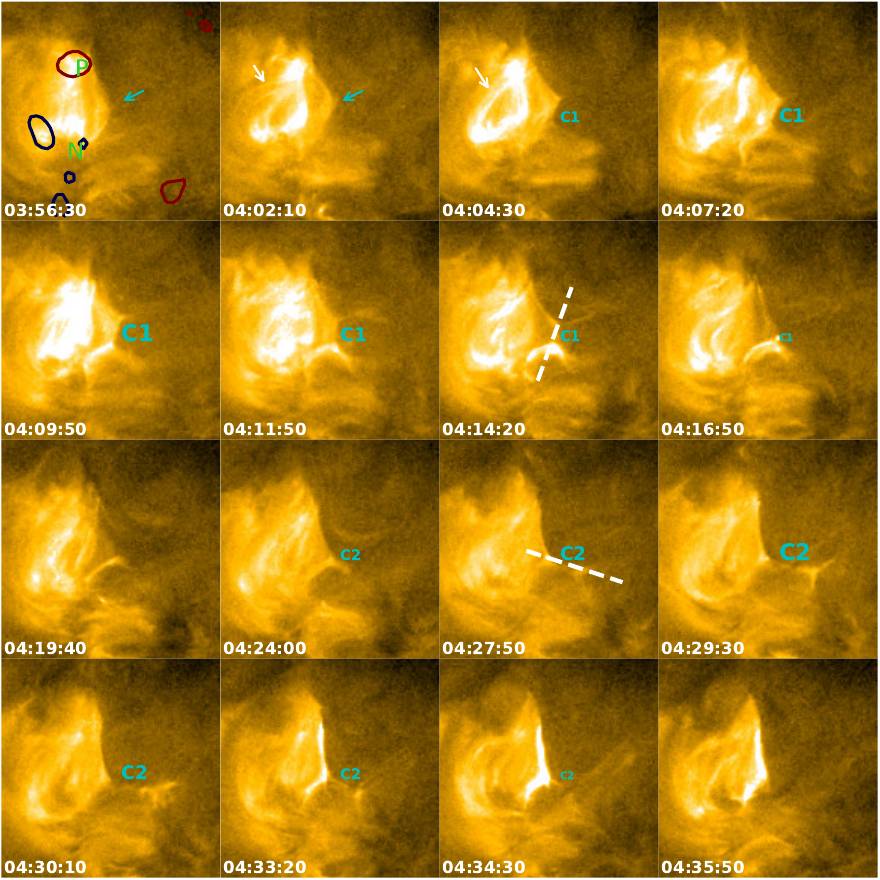}
\caption{Temporal evolution of the CBP and the associated current sheet  as seen in EUI 174 \AA\ images. Each panel has the same FOV as those in Figure2.  The overlaid red (blue) contours in the first panel reflect the simultaneous HMI magnetogram at levels of 30 (-30) G. The cyan arrows in the first two panels indicate the western loop of the CBP  expanding outward. The white arrows in the second and third panels indicate the rising and expanding loop-like structure from the core of the CBP. C1 and C2 represent the interest current sheets formed at the west edge of the CBP. Different size of the letters C1 and C2 indicates that the current sheets get longer or shorter. The dash lines show the different orientation of  C1 and C2. (An animation associated with this figure is available online)}
\end{figure}

\begin{figure}[ht!]
\plotone{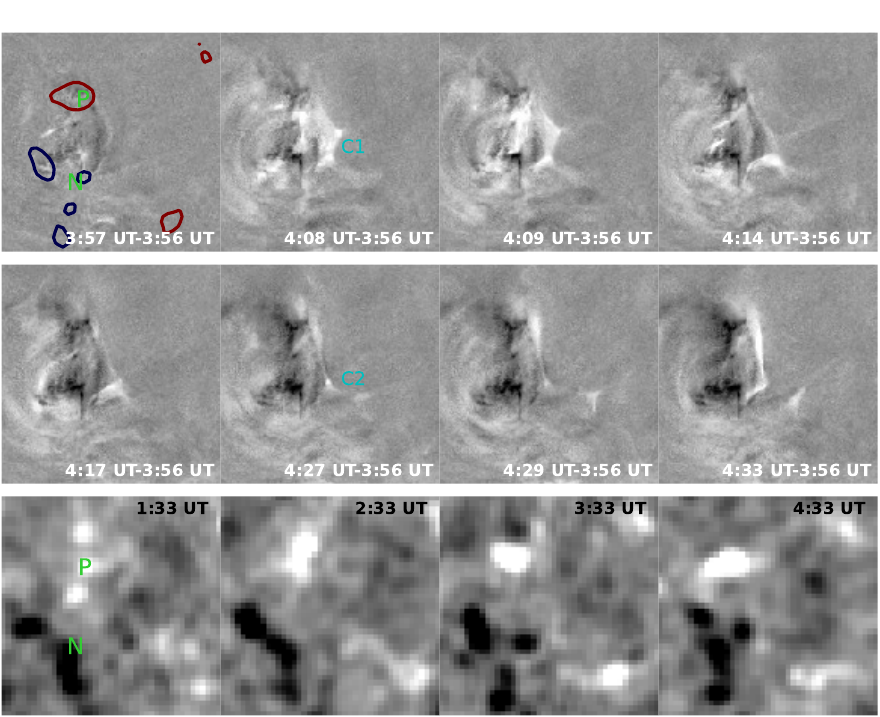}
\caption{{\it The first two rows:} The evolution of C1/C2 in consecutive base difference 174 \AA\ images. The  red (blue) contours have the same meaning as those in the previous figure.  {\it The last row:} HMI magnetograms showing the temporal evolution of the underlying magnetic poles before and during the formation of C1/C2. Each panel has the same FOV as the previous Figure. (An animation associated with this figure is available online)}
\end{figure}

\begin{figure}[ht!]
\plotone{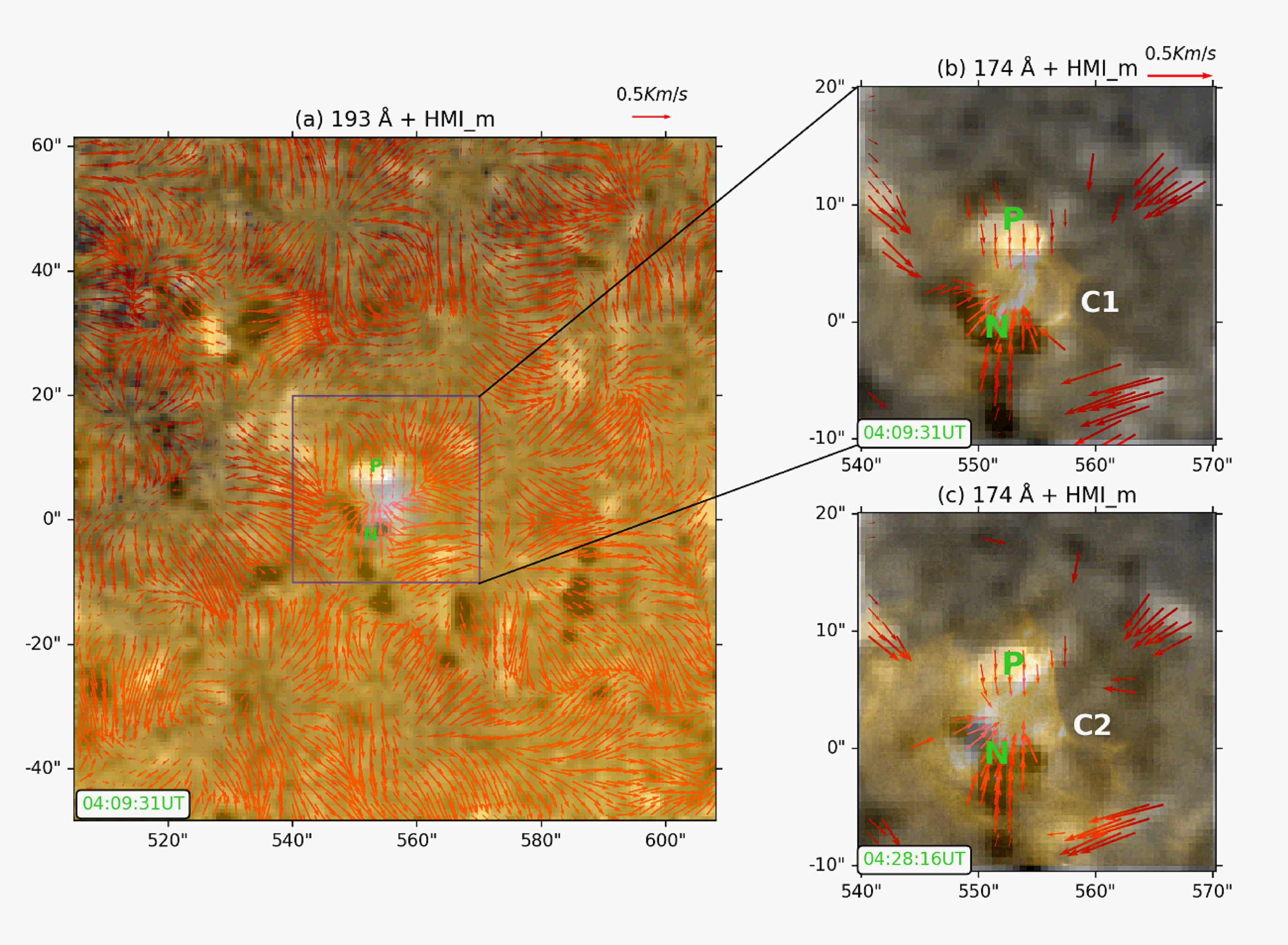}
\caption{{\it Panel a}: Combine image of  the simultaneous AIA 193 \AA\ observation and HMI magnetogram, superimposed with the  horizontal flow field (shown by red arrows) on photosphere. The horizontal flow field shows an averaged photospheric flow over 03:29-04:29 UT, which  is obtained by using the optical flow method (cv2.calcOpticalFlowFarneback in OpenCV) to calculate the movement of each pixel on the consecutive HMI magnetograms.  {\it Panel b/c}: show the zoomed-in region of the blue box, but combine the simultaneous EUI 174 \AA\ observations and HMI magnetograms to display C1/C2 associated with the underly photosphere activity..}
\end{figure}

\begin{figure}[ht!]
\epsscale{1.}
\plotone{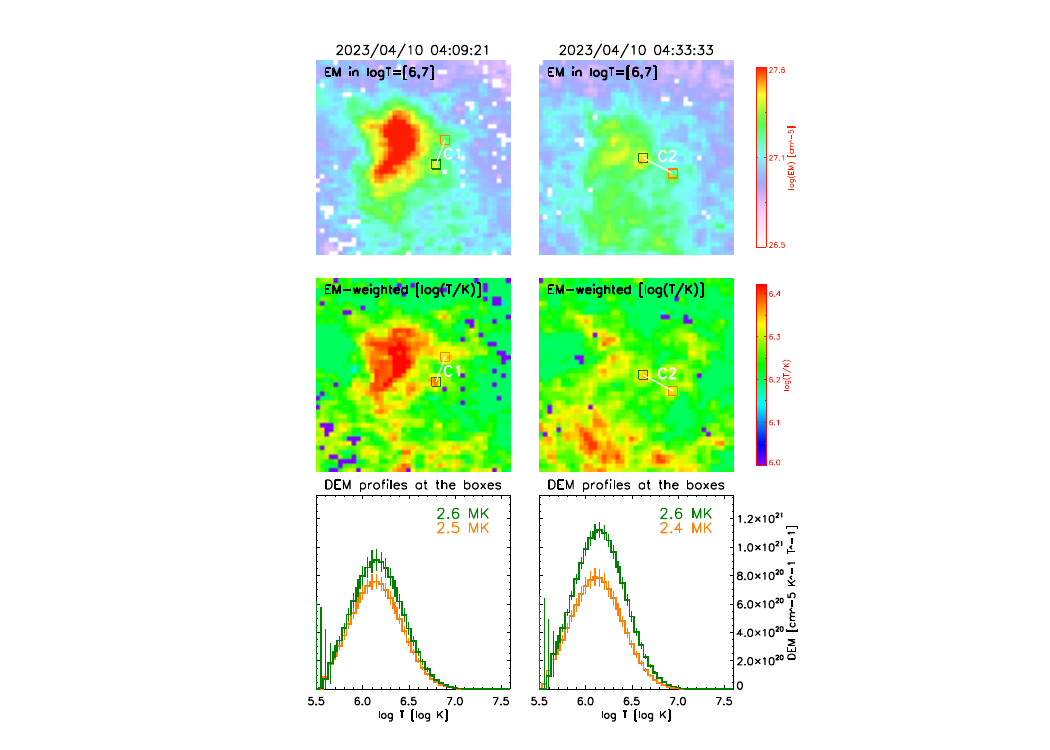}
\caption{DEM results inversioned from AIA data at two time points when C1/C2 appeared respectively. The derived EM maps (first row) and EM-weighted temperature (second row) maps have the same FOV as those panels in Figure 2-4.   The white lines indicate the locations of C1/C2 roughly by on the EM maps and EM-weighted temperature maps, while the green/orange boxes mark each end of C1/C2. The green/orange DEM profiles at the bottom row are extracted from the averaged DEM distributions over 9 pixels within the green/orange boxes. The error bars along the DEM curves reflect the standard deviation of DEM calculations from 100 Monte Carlo simulations by adding AIA data noise into the DEM inversions. The present temperature values are the DEM-weighted temperatures that are calculated from each DEM profile base on the definition $T={\int}{\mbox{DEM}}(T)TdT/{\int}{\mbox{DEM}}(T)dT$. }
\end{figure}

\begin{figure}[ht!]
\epsscale{1.}
\plotone{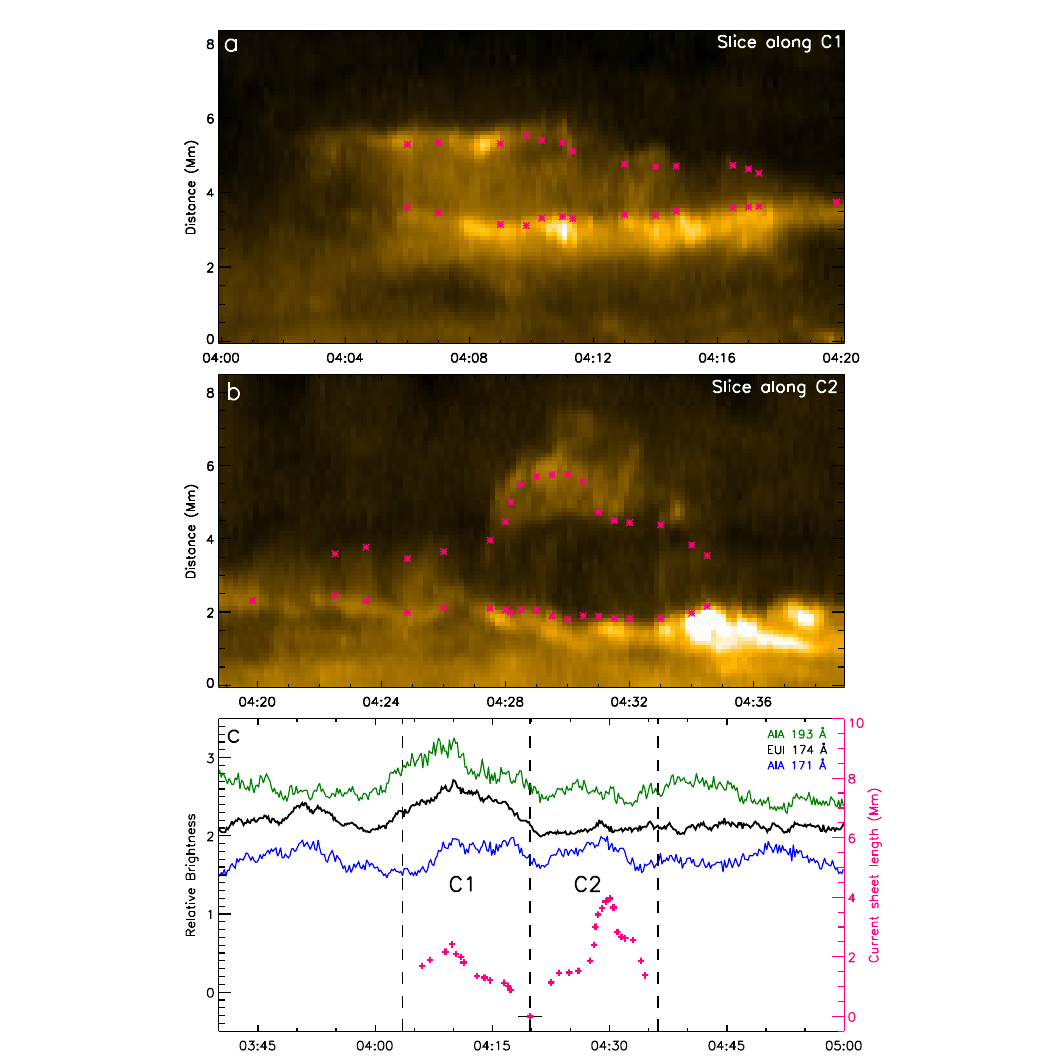}
\caption{{\it Panel a/b:} 174 \AA\ time-slice plots that is made from the slices along C1/C2 (registered by the long dashed lines in Fig.3). The overlaid pink asterisk indicate  the position of the ends of C1/C2 at different moment. {\it Panel c:} The blue, green and dark light curves show temporal evolution of the CBP brightness relative to the average background intensity in AIA 171 \AA\ , AIA 193 \AA\ and EUI 174 \AA\ . The vertical dashed lines mark the duration of C1 and C2. The pink plus profile indicates the temporal evolution of the length of C1/C2. The horizontal bar at the transition point from C1 to C2 indicates a time interval (~3 mins) when the current sheet length is close to 0 during the transition.}
\end{figure}


\bibliography{sample631}{}

\begin{thebibliography}{}
\bibitem[Alipour \& Safari(2015)]{Alipour2015} Alipour, N. \& Safari, H.\ 2015, \apj, 807, 175. doi:10.1088/0004-637X/807/2/175
\bibitem[Aschwanden \& Boerner(2011)]{Aschwanden2011} Aschwanden, M.~J. \& Boerner, P.\ 2011, \apj, 732, 2, 81. doi:10.1088/0004-637X/732/2/81
\bibitem[Brown et al.(2001)]{Brown2001} Brown, D.~S., Parnell, C.~E., Deluca, E.~E., Golub, L., \& McMullen, R.~A.\ 2001, \solphys, 201, 305
\bibitem[Cai et al.(2024)]{Cai2024} Cai, Q., Li, H., Wang, J., et al.\ 2024, \aap, 691, A309. doi:10.1051/0004-6361/202449396
\bibitem[Chen et al.(1999)]{Chen1999} Chen, P.~F., Fang, C., Tang, Y.~H., et al.\ 1999, \apj, 513, 516. doi:10.1086/306823
\bibitem[Cheng et al.(2023)]{Cheng2023} Cheng, X., Priest, E.~R., Li, H.~T., et al.\ 2023, Nature Communications, 14, 2372. doi:10.1038/s41467-023-38149-6
\bibitem[Cheung et al.(2015)]{Cheung2015} Cheung, M.~C.~M., Boerner, P., Schrijver, C.~J., et al.\ 2015, \apj, 807, 143. doi:10.1088/0004-637X/807/2/143
\bibitem[Craig \& McClymont(1991)]{Craig1991} Craig, I.~J.~D. \& McClymont, A.~N.\ 1991, \apjl, 371, L41. doi:10.1086/185997
\bibitem[Ding et al.(2024)]{Ding2024} Ding, T., Zhang, J., Fang, Y., et al.\ 2024, \apj, 964, 58. doi:10.3847/1538-4357/ad2683
\bibitem[Duan et al.(2024)]{Duan2024} Duan, Y., Tian, H., Chen, H., et al.\ 2024, \apjl, 962, L38. doi:10.3847/2041-8213/ad24f3
\bibitem[Golub et al.(1974)]{Golub1974} Golub, L., Krieger, A.~S., Silk, J.~K., et al.\ 1974, \apjl, 189, L93. doi:10.1086/181472
\bibitem[Golub et al.(1976)]{Golub1976} Golub, L., Krieger, A.~S., \& Vaiana, G.~S.\ 1976, \solphys, 50, 311
\bibitem[Goff et al.(2007)]{Goff2007} Goff, C.~P., van Driel-Gesztelyi, L., D{\'e}moulin, P., et al.\ 2007, \solphys, 240, 283. doi:10.1007/s11207-007-0260-4
\bibitem[Joshi et al.(2020)]{Joshi2020} Joshi, R., Chandra, R., Schmieder, B., et al.\ 2020, \aap, 639, A22. doi:10.1051/0004-6361/202037806
\bibitem[Habbal \& Withbroe(1981)]{Habbal1981} Habbal, S.~R. \& Withbroe, G.~L.\ 1981, \solphys, 69, 77. doi:10.1007/BF00151257
\bibitem[Habbal \& Harvey(1988)]{Habbal1988} Habbal, S.~R. \& Harvey, K.~L.\ 1988, \apj, 326, 988. doi:10.1086/166156
\bibitem[Habbal et al.(1990)]{Habbal1990} Habbal, S.~R., Dowdy, J.~F., \& Withbroe, G.~L.\ 1990, \apj, 352, 333. doi:10.1086/168540
\bibitem[Harvey et al.(1993)]{Harvey1993} Harvey, K.~L., Strong, K.~T., Nitta, N., et al.\ 1993, Advances in Space Research, 13, 27. doi:10.1016/0273-1177(93)90453-I
\bibitem[Honarbakhsh et al.(2016)]{Honarbakhsh2016} Honarbakhsh, L., Alipour, N., \& Safari, H.\ 2016, \solphys, 291, 3, 941. doi:10.1007/s11207-016-0858-5
\bibitem[Hong et al.(2014)]{Hong2014} Hong, J., Jiang, Y., Yang,  J., et al.\ 2014, \apj, 796, 73 
\bibitem[Hong et al.(2019)]{Hong2019} Hong, J., Yang, J., Chen, H., et al.\ 2019, \apj, 874, 146. doi:10.3847/1538-4357/ab0c9d
\bibitem[Hosseini Rad et al.(2021)]{Hosseini2021} Hosseini Rad, S., Alipour, N., \& Safari, H.\ 2021, \apj, 906, 1, 59. doi:10.3847/1538-4357/abc8e8
\bibitem[Huang et al.(2012)]{Huang2012} Huang, Z., Madjarska, M.~S., Doyle, J.~G., \& Lamb, D.~A.\ 2012, \aap, 548, A62
\bibitem[Innes et al.(2009)]{Innes2009} Innes, D.~E., Genetelli, A., Attie, R., et al.\ 2009, \aap, 495, 1, 319. doi:10.1051/0004-6361:200811011
\bibitem[Lemen et al.(2012)]{Lemen2012} Lemen, J.~R., Title, A.~M., Akin, D.~J., et al.\ 2012, \solphys, 275, 17
\bibitem[Li et al.(2021)]{Li2021} Li, L., Peter, H., Chitta, L.~P., et al.\ 2021, \apj, 908, 213. doi:10.3847/1538-4357/abd47e
\bibitem[Li et al.(2020)]{Li2020} Li, D., Feng, S., Su, W., et al.\ 2020, \aap, 639, L5. doi:10.1051/0004-6361/202038398
\bibitem[Li(2022)]{Li2022} Li, D.\ 2022, \aap, 662, A7. doi:10.1051/0004-6361/202142884
\bibitem[Liu et al.(2016)]{Liu2016} Liu, R., Chen, J., Wang, Y., et al.\ 2016, Scientific Reports, 6, 34021. doi:10.1038/srep34021
\bibitem[Longcope(1998)]{Longcope1998} Longcope, D.~W.\ 1998, \apj, 507, 433
\bibitem[Longcope et al.(2001)]{Longcope2001} Longcope, D. W., Kankelborg, C. C., Nelson, J. L., \& Pevtsov, A. A.\ 2001, \apj, 553, 429
\bibitem[Madjarska(2019)]{Madjarska2019} Madjarska, M.~S.\ 2019, Living Reviews in Solar Physics, 16, 2. doi:10.1007/s41116-019-0018-8
\bibitem[Madjarska et al.(2020)]{Madjarska2020} Madjarska, M.~S., Galsgaard, K., Mackay, D.~H., et al.\ 2020, \aap, 643, A19. doi:10.1051/0004-6361/202038287
\bibitem[McIntosh(2007)]{McIntosh2007} McIntosh, S.~W.\ 2007, \apj, 670, 2, 1401. doi:10.1086/521948
\bibitem[Mou et al.(2018)]{Mou2018} Mou, C., Madjarska, M.~S., Galsgaard, K., et al.\ 2018, \aap, 619, A55. doi:10.1051/0004-6361/201833243
\bibitem[McLaughlin et al.(2012)]{McLauhlin2012} McLaughlin, J.~A., Thurgood, J.~O., \& MacTaggart, D.\ 2012, \aap, 548, A98. doi:10.1051/0004-6361/201220234
\bibitem[Murray et al.(2009)]{Murray2009} Murray, M.~J., van Driel-Gesztelyi, L., \& Baker, D.\ 2009, \aap, 494, 329 
\bibitem[M{\"u}ller et al.(2020)]{Muller2020} M{\"u}ller, D., St. Cyr, O.~C., Zouganelis, I., et al.\ 2020, \aap, 642, A1. doi:10.1051/0004-6361/202038467
\bibitem[Ni et al.(2017)]{Ni2017} Ni, L., Zhang, Q.-M., Murphy, N.~A., et al.\ 2017, \apj, 841, 1, 27. doi:10.3847/1538-4357/aa6ffe
\bibitem[N{\'o}brega-Siverio et al.(2023)]{Nobrega2023} N{\'o}brega-Siverio, D., Moreno-Insertis, F., Galsgaard, K., et al.\ 2023, \apjl, 958, L38. doi:10.3847/2041-8213/ad0df0
\bibitem[Ning \& Guo(2014)]{Ning2014} Ning, Z. \& Guo, Y.\ 2014, \apj, 794, 79. doi:10.1088/0004-637X/794/1/79
\bibitem[Ning et al.(2020)]{Ning2020} Ning, Z.-J., Li, D., \& Zhang, Q.-M.\ 2020, Research in Astronomy and Astrophysics, 20, 138. doi:10.1088/1674-4527/20/9/138
\bibitem[Karampelas et al.(2022)]{Karampelas2022a} Karampelas, K., McLaughlin, J.~A., Botha, G.~J.~J., et al.\ 2022, \apj, 925, 195. doi:10.3847/1538-4357/ac3b53
\bibitem[Karampelas et al.(2023)]{Karampelas2023} Karampelas, K., McLaughlin, J.~A., Botha, G.~J.~J., et al.\ 2023, \apj, 943, 131. doi:10.3847/1538-4357/acac90
\bibitem[Kariyappa \& Varghese(2008)]{Kariyappa2008} Kariyappa, R., \& Varghese, B. A.\ 2008, \aap, 485, 289
\bibitem[Kariyappa et al.(2011)]{Kariyappa2011} Kariyappa, R., DeLuca, E. E., Saar, S. H., Golub, L., DamšŠ, L., Pevtsov, A. A., \& Varghese, B. A.\ 2010, \aap, 526, A78
\bibitem[Kumar et al.(2011)]{Kumar2011} Kumar, M., Srivastava, A.~K., \& Dwivedi, B.~N.\ 2011, \mnras, 415, 1419. doi:10.1111/j.1365-2966.2011.18792.x
\bibitem[Kupriyanova et al.(2020)]{Kupriyanova2020} Kupriyanova, E., Kolotkov, D., Nakariakov, V., et al.\ 2020, Solar-Terrestrial Physics, 6, 3. doi:10.12737/stp-61202001
\bibitem[Kwon et al.(2012)]{Kwon2012} Kwon, R.-Y., Chae, J., Davila, J.~M., et al.\ 2012, \apj, 757, 167
\bibitem[Ramsey et al.(2023)]{Ramsey2023} Ramsey, B., Verwichte, E., \& Morgan, H.\ 2023, \aap, 679, A10. doi:10.1051/0004-6361/202346757
\bibitem[Rochus et al.(2020)]{Rochus2020} Rochus, P., Auch{\`e}re, F., Berghmans, D., et al.\ 2020, \aap, 642, A8. doi:10.1051/0004-6361/201936663
\bibitem[Samanta et al.(2015)]{Samanta2015} Samanta, T., Banerjee, D., \& Tian, H.\ 2015, \apj, 806, 172. doi:10.1088/0004-637X/806/2/172
\bibitem[Schou et al.(2012)]{Schou2012} Schou, J., Scherrer, P.~H., Bush, R.~I., et al.\ 2012, \solphys, 275, 229
\bibitem[Shen et al.(2022)]{Shen2022} Shen, Y., Yao, S., Tang, Z., et al.\ 2022, \aap, 665, A51. doi:10.1051/0004-6361/202243924
\bibitem[Sterling \& Moore(2001)]{Sterling2001} Sterling, A.~C. \& Moore, R.~L.\ 2001, \jgr, 106, 25227. doi:10.1029/2000JA004001
\bibitem[Strong et al.(1992)]{Strong1992} Strong, K. T., Harvey, K., Hirayama, T., et al.\ 1992, \pasj, 44, L161
\bibitem[Sweet(1958)]{1958Sweet} Sweet, P.~A.\ 1958, Electromagnetic Phenomena in Cosmical Physics, 6, 123
\bibitem[Talbot et al.(2024)]{Tabaot2024} Talbot, J., McLaughlin, J.~A., Botha, G.~J.~J., et al.\ 2024, \apj, 965, 133. doi:10.3847/1538-4357/ad2a5d
\bibitem[Tian et al.(2008)]{Tian2008} Tian, H., Xia, L.-D., \& Li, S.\ 2008, \aap, 489, 741. doi:10.1051/0004-6361:200810146
\bibitem[O'Dwyer et al.(2010)]{Odwyer2010} O'Dwyer, B., Del Zanna, G., Mason, H.~E., et al.\ 2010, \aap, 521, A21. doi:10.1051/0004-6361/201014872
\bibitem[Parker(1957)]{Parker1957} Parker, E.~N.\ 1957, \jgr, 62, 509
\bibitem[Pesnell et al.(2012)]{Pesnell2011} Pesnell, W.~D., Thompson, B.~J., \& Chamberlin, P.~C.\ 2012, \solphys, 275, 3 
\bibitem[Petschek(1964)]{Petschek1964} Petschek, H.~E.\ 1964, NASA Special Publication, 50, 425 
\bibitem[Priest et al.(1994)]{Priest1994} Priest, E.~R., Parnell, C.~E., \& Martin, S.~F.\ 1994, \apj, 427, 459
\bibitem[Ugarte-Urra et al.(2004)]{Ugarte2004} Ugarte-Urra, I., Doyle, J.~G., Madjarska, M.~S., \& O'Shea, E.\ 2004, \aap, 418, 313
\bibitem[Vaiana et al.(1973)]{Vaiana1973} Vaiana, G.~S., Davis,J.~M., Giacconi, R., et al.\ 1973, \apjl, 185, L47
\bibitem[Webb et al.(1993)]{Webb1993} Webb, D.~F., Martin, S.~F., Moses, D., et al.\ 1993, \solphys, 144, 15. doi:10.1007/BF00667979
\bibitem[Wyper et al.(2018)]{Wyper2018} Wyper, P.~F., DeVore, C.~R., Karpen, J.~T., et al.\ 2018, \apj, 864, 165. doi:10.3847/1538-4357/aad9f7
\bibitem[Xu et al.(2024)]{Xu2024} Xu, Z., Yan, X., Yang, L., et al.\ 2024, \mnras, 530, 1, 473. doi:10.1093/mnras/stae822
\bibitem[Xue et al.(2019)]{Xue2019} Xue, Z., Yan, X., Jin, C., et al.\ 2019, \apjl, 874, L27. doi:10.3847/2041-8213/ab1135
\bibitem[Yan et al.(2022)]{Yan2022} Yan, X., Xue, Z., Jiang, C., et al.\ 2022, Nature Communications, 13, 640. doi:10.1038/s41467-022-28269-w
\bibitem[Yang et al.(2015)]{Yang2015} Yang, S., Zhang, J., \& Xiang, Y.\ 2015, \apjl, 798, L11. doi:10.1088/2041-8205/798/1/L11
\bibitem[Yang et al.(2018)]{Yang2018} Yang, B., Yang, J., Bi, Y., et al.\ 2018, \apj, 861, 135. doi:10.3847/1538-4357/aac37f
\bibitem[Yang et al.(2023)]{Yang2023} Yang, J., Hong, J., Yang, B., et al.\ 2023, \apj, 942, 86. doi:10.3847/1538-4357/aca66f
\bibitem[Yang et al.(2024)]{Yang2024} Yang, L., Yan, X., Xue, Z., et al.\ 2024, \mnras, 528, 1094. doi:10.1093/mnras/stad3876
\bibitem[Zhou et al.(2024)]{Zhou2024} Zhou, X., Shen, Y., Zhou, C., et al.\ 2024, Science China Physics, Mechanics, and Astronomy, 67, 259611. doi:10.1007/s11433-023-2309-5
\bibitem[Zhang et al.(2001)]{Zhang2001} Zhang, J., Kundu, M. R., \& White, S. M.\ 2001, \solphys, 198, 347
\bibitem[Zhang et al.(2012)]{Zhang2012} Zhang, Q. M., Chen, P. F., Guo, Y., Fang, C., \& Ding, M. D.\ 2012, \apj, 746, 19
\bibitem[Zhang et al.(2014)]{Zhang2014} Zhang, Q.~M., Chen, P.~F., Ding, M.~D., \& Ji, H.~S.\ 2014, \aap, 568, A30 
\end{thebibliography}
\bibliographystyle{aasjournal}



\end{document}